\documentclass[a4paper,11pt]{article}
\usepackage{pos}
\usepackage{booktabs}
\usepackage{adjustbox}
\usepackage{subcaption}
\usepackage{cleveref}

\usepackage{mathrsfs}
\usepackage{mathtools}
\usepackage{titlesec}
\usepackage[utf8]{inputenc}
\usepackage{graphicx}
\usepackage{psfrag,epsf}
\usepackage{slashed}
\usepackage{enumitem}
\usepackage{epigraph}
\usepackage{siunitx}
\usepackage{setspace}
\usepackage{bbold}
\usepackage{tabstackengine}

\DeclareMathOperator{\Tr}{tr}
\title{Deconfinement critical point of heavy quark effective lattice theories}

\author[a]{Jangho Kim}
\author*[b]{Anh Quang Pham}
\author[b]{Owe Philipsen}

\affiliation[a]{Institute for Advanced Simulation (IAS-4),\\
Forschungszentrum Jülich, Wilhelm-Johnen-Straße, 52428 Jülich, Germany}

\affiliation[b]{Institut für Theoretische Physik, Goethe-Universität Frankfurt am Main,\\
Max-von-Laue-Str. 1, 60438 Frankfurt am Main, Germany}

\emailAdd{j.kim@fz-juelich.de, pham, philipsen@itp.uni-frankfurt.de}

\abstract{Effective three-dimensional Polyakov loop theories derived from QCD by strong coupling and hopping expansions are valid for heavy quarks and can also be applied to finite chemical potential $\mu$, due to their considerably milder sign problem. We apply the Monte-Carlo method to the $N_f=1,2$ effective theories up to $\mathcal{O}(\kappa^4)$ in the hopping parameter at $\mu=0$ to determine the critical quark mass, at which the first-order deconfinement phase transition terminates. The critical end point obtained from the effective theory to order $\mathcal{O}(\kappa^2)$ agrees well with 4-dimensional QCD simulations with a hopping expanded determinant by the WHOT-QCD collaboration. We also compare with full QCD simulations and thus obtain a measure for the validity of both the strong coupling and the hopping expansion in this regime.}

\FullConference{%
 The 38th International Symposium on Lattice Field Theory, LATTICE2021
  26th-30th July, 2021
  Zoom/Gather@Massachusetts Institute of Technology
}

\begin{document}
\maketitle

\section{Introduction}

Significant effort in theory and experiment is dedicated to explore QCD under extreme conditions, 
such as high temperature ($T$) or baryon chemical potential ($\mu_B=3\mu$). 
Despite considerable progress for zero and small chemical potential, 
the QCD phase diagram is still largely unknown, because the fermion sign problem prohibits Monte Carlo 
investigations with finite chemical potential, especially in the cold and dense region, which is central to astrophysics.

These difficulties have motivated the development of effective theories derived from lattice QCD by strong coupling and hopping parameter expansions, which are vaild in the heavy quark regime. 
These can either be solved by fully analytic series expansion techniques known 
from statistical mechanics \cite{Kim:2020atu, Kim:2019ykj, Glesaaen:2015vtp}, or by numerical simulations if their 
sign problem is sufficiently mild~\cite{Langelage:2014vpa, Scior:2015vra}. 

 For the $SU(3)$ pure gauge theory, the deconfinement transition spontaneously breaks the global $Z_3$ center symmetry, and  is of 
 first order. The presence of dynamical quarks breaks the center symmetry explicitly, and leads to a weakening of the  deconfinement transition with decreasing quark mass until it vanishes at a critical point in the 3D Ising universality class. 
 For still lighter quark masses, the transition becomes an analytic crossover.
This behavior is also inherited by the effective theory. 

Here we update previous studies of three-dimensional effective theories at zero chemical potential in the heavy quark regime using standard Monte Carlo simulations. 
We calculate the deconfinement phase transition and its critical end point for $N_f=1,2$, for $N_t=4,6$ for different
truncations of the three-dimensional effective theory up to order $\mathcal{O}(\kappa^4)$ in the hopping parameter. Our results 
are compared with 
simulations of four-dimensional QCD after hopping expansion \cite{Ejiri:2019csa}, as well
as with full QCD simulations \cite{Cuteri:2020yke}. 

\section{3D Effective lattice theories}

First we briefly discuss the derivation of the effective action. For more detail, see for instance \cite{Langelage:2014vpa,Fromm:2011qi}. The starting point is a $(3+1)$-dimensional lattice with Wilson gauge and fermion action for $N_f$ flavors. Integration over the fermion fields leads to a partition function of the form
\begin{align}
	Z=\int[\mathrm{d} U_\mu]\exp[-S_g]\prod_{f=1}^{N_f}\det[D_f], \hspace{1cm}S_g=-\frac{\beta}{2N_c}\sum_p\left[\Tr U_p+\Tr U^\dagger_p\right],
\end{align}
where $\det [D_f]$ is the fermion determinant and $S_g$ is the Wilson gauge action. 
The effective theory arises after integrating over the spatial link variables, 
\begin{align}
Z=\int [\mathrm{d}U_0]\exp[-S_{\text{eff}}], \hspace{0.5cm}\exp[-S_{\text{eff}}] =\int	 [\mathrm{d}U_i]\exp[-S_g]\prod_{f=1}^{N_f}\det [D_f].
\end{align}
The effective action depends on temporal Wilson lines $W_{\vec{x}}$ only, whose traces are the Polyakov loops
\begin{align}
\label{eq:polyakov_loop}
L_{\vec{x}}=\Tr W_{\vec{x}}=\Tr \prod_{t=0}^{N_t-1}U_0(\vec{x},t).	
\end{align}
Our investigation is based on truncated effective theories, where the spatial link integration is performed after a combined 
character (i.e.~effectively resummed strong coupling) and hopping parameter expansion.
For the Yang-Mills part, to leading order we obtain a nearest-neighbor two-point interaction
\begin{align}
S_{\text{eff}}^g&=-\sum_{\langle\vec{x},\vec{y}\rangle}\log\left[1+\lambda(u,N_t)(L_{\vec{x}}L^*_{\vec{y}}+L^*_{\vec{x}}L_{\vec{y}})\right],
\label{eq:pure_gauge_effective_theory}
\end{align}
where the effective coupling reads
\begin{align}
\lambda(u,N_t=4)&=u^{4}\exp\left[4(4u^4+12u^5-14u^6-36u^7+\frac{295}{2}u^8+\frac{
1851}{10}u^9+\frac{1035317}{5120}u^{10})\right],\\
\lambda(u,N_t\geq 6)&=u^{N_t}\exp\left[N_t(4u^4+12u^5-14u^6-36u^7+\frac{295}{2}u^8+\frac{
1851}{10}u^9+\frac{1055797}{5120}u^{10})\right].
\end{align}
Here
$u(\beta)=\beta/18+\beta^2/216+\ldots \in[0,1]$,  is
the coefficient of the fundamental representation character.  It is a numerically 
known function over the entire range of lattice gauge couplings 
and constitutes the expansion parameter. 
Higher order interaction terms can be found in \cite{Langelage:2010yr}. 

The contributions of the fermion determinant are computed by a hopping expansion. It is factored 
into temporal and spatial hops in forward and backward directions, $T=T^++T^-$ and $S=S^++S^-$, 
\begin{align}
\det [D_f]=\det[1-T-S]=\det[1-T]\det[1-(1-T)^{-1}S]\nonumber
=\det[D_{\text{stat}}]\det[D_{\text{kin}}].
\label{eq:stat_kin_determinant}
\end{align}
To leading order the fermion determinant represents static quarks with only temporal hops, and can be reformulated in terms of Polyakov loops, 
\begin{equation}
\label{eq:stat_quark_action1}
S_0=-\log\det[D_{\text{stat}}]=-\log\left(\prod_{\vec{x}}[1+h_1L_{\vec{x}}+h_1^2L^*_{\vec{x}}+h_1^3]^2[1+\bar{h}_1L^*_{\vec{x}}+\bar{h}^2_1L_{\vec{x}}+\bar{h}_1^3]^2\right).
\end{equation}
To leading order in the combined expansions, the coefficients are the heavy quark fugacities,
\begin{align}
h_1(\mu)=(2\kappa e^{a\mu})^{N_t}=e^{\frac{\mu-m_q}{T}}=\bar{h}_1(-\mu).
\end{align}
In the strong coupling limit $\beta=0$ the constituent quark mass is given by $am_q=-\log(2\kappa)$ \cite{Green:1983sd}.
The kinetic quark determinant is evaluated by further splitting $D_{\text{kin}}$ into parts from positive and negative spatial hops, $P=\sum_i P_i=(1-T)^{-1}S^+$ and $M=\sum_i M_i=(1-T)^{-1}S^-$. Here the
static quark propagator $Q_{\text{stat}}^{-1}=(1-T)^{-1}$ enters, which is known to all orders in the hopping parameter.
Then the expansion of the kinetic quark determinant proceeds as
\begin{align}
\det [D_{\text{kin}}]&=\det[1-P-M]=\exp[\Tr \log(1-P-M)]\nonumber\\
&=\exp\left[-\Tr PM-\Tr PPMM-\frac{1}{2}\Tr PMPM +\mathcal{O}(\kappa^6)\right]\nonumber\\
&=1-\Tr PM - \Tr PPMM - \frac{1}{2} \Tr PMPM +\frac{1}{2}(\Tr PM)^2 +\mathcal{O}(\kappa^6).
\label{eq:expand_exp_to_do_spatial_integral}
\end{align} 
In order to perform the spatial gauge integrals on $\det [D_{\text{kin}}]$, it is necessary to expand down the exponential as shown in equation \eqref{eq:expand_exp_to_do_spatial_integral}. 
Further details of the derivation up to order $\mathcal{O}(\kappa^4)$ can be found in \cite{Langelage:2014vpa,Neuman2015zb}.
The process of resummation is employed to obtain the exponential expression of the effective theory which then improves the convergence of the effective theory, since it includes an infinite number of higher-order graphs. Finally, the leading term of the kinetic determinant contributes to a two-point interaction and is of order $\kappa^2$, 
\begin{align}
-S_2&=-\int[\mathrm{d}U_j]\sum_i\Tr P_iM_i=\sum_i\int[\mathrm{d}U_j]\Tr\left[Q_{\text{stat}}^{-1}S^+_i Q^{-1}_{\text{stat}}S^-_i\right]\nonumber\\
&=-2h_2\sum_{i,\vec{x}}\left[\left(\Tr\frac{h_1W_{\vec{x}}}{1+h_1W_{\vec{x}}} - \Tr\frac{\bar{h}_1W^\dagger_{\vec{x}}}{1+\bar{h}_1W^\dagger_{\vec{x}}} \right)\left(\Tr\frac{h_1W_{\vec{x}+\hat{i}}}{1+h_1W_{\vec{x}+\hat{i}}} - \Tr\frac{\bar{h}_1W^\dagger_{\vec{x}+\hat{i}}}{1+\bar{h}_1W^\dagger_{\vec{x}+\hat{i}}} \right)\right].
\label{eq:kappa2_term}
\end{align}
To re-express the trace over a rational function containing temporal Wilson lines in terms of Polyakov loops, we use the generating function
\begin{align}
G(\alpha,\beta)=\log\det(\alpha+\beta h_1 W),
\end{align}
from which 
\begin{align}
W_{nm}=\Tr\frac{(h_1W)^m}{(1+h_1W)^n}=\frac{(-1)^{n-1}}{(n-1)!}\frac{\partial^{n-m}}{\partial\alpha^{n-m}}\frac{\partial^m}{\partial\beta^m}G(\alpha,\beta)\bigg|_{\alpha=\beta=1}.
\end{align}
The contribution to order $\kappa^4$ is 
\begin{equation}
-S_4=\int[\mathrm{d}U_j] \Big(- \Tr PPMM - \frac{1}{2} \Tr PMPM +\frac{1}{2}(\Tr PM)^2\Big)
\end{equation}
and after complete evaluation is too long to be printed here. For explicit expressions, see \cite{Langelage:2014vpa,Neuman2015zb}.
The effective action to $\mathcal{O}(\kappa^4)$ used in our simulations is then
\begin{align}
S_{\text{eff}}&=S_{\text{eff}}^g+S_0+S_2+S_4.
\label{eq:full_action}
\end{align}
Note that moving away from the strong coupling limit leads to corrections to the fermion couplings $h_i$ coming from mixed graphs with contributions from non-vanishing gauge coupling as well. 
Our effective couplings including corrections for $N_t \geq 4$ are
\begin{align}
h_1(u,\kappa,N_t)&=(2\kappa e^{a\mu})^{N_t}\exp\left[6N_t\kappa^2u\left(\frac{1-u^{N_t-1}}{1-u}+4u^4-12\kappa^2+9\kappa^2u+4\kappa^2u^2-4\kappa^4\right)\right],\\
h_2(u,\kappa,N_t)&=\frac{\kappa^2N_t}{N_c}\left(1+2\frac{u-u^{N_t}}{1-u}+8u^5+16\kappa^2u^4\right),\\
h_3^1(u,\kappa,N_t)&=\frac{N_t(N_t-1)\kappa^4}{N_c^2}\left[1+\frac{8}{3}(u+u^2+4u^5+8\kappa^3u^4)\right], \hspace{0.5cm} \text{for} \hspace{0.5cm} N_t=4,\\
h_3^2(u,\kappa,N_t)&=\frac{\kappa^4N_t}{N_c^2}\left[1+4\frac{u-u^{N_t}}{1-u}+16u^5+32\kappa^3u^4\right],\\
h_3^3(u,\kappa,N_t)&=\frac{\kappa^4N_t^2}{N_c^2}\left[1+4\frac{(1-u^{N_t})(u-u^{N_t})}{(1-u)^2}+16u^5+32\kappa^3u^4\right],\\
h_3^4(u,\kappa,N_t)&=\frac{\kappa^4 uN_t}{2N_c^3}\left[1+4u^4+16\kappa^3u^4\right],
\end{align}
where $h_3^1, \cdots, h_3^4$ are effective couplings of $S_4$ coupled to $\kappa^4$ graphs with different gauge corrections.
\section{Numerical results}
Since the effective theory depends only on the Polyakov loops, the numerical investigation for the effective theory can be performed directly with Metropolis updates of the temporal links, which live on a three-dimensional lattice.
The bare fermion mass $am$ is controlled via the hopping parameter $\kappa=(2(am+4))^{-1}$. Finite temperature on the lattice is given by the inverse temporal extent of the original lattice, $T=1/a(\beta)N_t$. We will work with $N_f=1,2$, and at fixed $N_t=4,6$ for the effective theory up to order $\mathcal{O}(\kappa^2)$, and at $N_t=4$ for the $\mathcal{O}(\kappa^4)$ effective theory. This work aims to map the phase structure of heavy QCD at zero chemical potential, i.e., in the $(u, \kappa)$ parameter space, for different approximations of the effective theory. 

The observable used in this work is the Polyakov loop, $\mathcal{O}\equiv |L|$, which is a true order parameter of QCD in the limit $m_q\rightarrow\infty$, and signals a phase transition. 
We then construct the susceptibility, the skewness and the kurtosis as 
\begin{align}
\chi=N_s^3(\langle\mathcal{O}^2\rangle-\langle\mathcal{O}\rangle^2),\, B_{3}=\frac{\langle(\mathcal{O}-\langle\mathcal{O}\rangle)^3\rangle}{\langle(\mathcal{O}-\langle\mathcal{O}\rangle)^2\rangle^{3/2}},\,B_{4}=\frac{\langle(\mathcal{O}-\langle\mathcal{O}\rangle)^4\rangle}{\langle(\mathcal{O}-\langle\mathcal{O}\rangle)^2\rangle^{2}}.
\end{align} 
The statistical error of these quantities are determined by a jackknife analysis.
\begin{table}[t]
  \begin{center}
    \begin{tabular}{ccccc}
    \toprule
       & \text{Crossover} & \text{first-order triple} & \text{Tricritical} & \text{3D Ising}\\
      \hline
     $B_4(\kappa_c,\infty)$ & 3 & 1.5 & 2 & 1.604 \\
     $\nu$ & - & 1/3 & 1/2 & 0.6301(4)\\
     $\gamma$ & - & 1 & 1 & 1.2372(5) \\
      \bottomrule
    \end{tabular}
    \caption{Critical values for $\nu$, $\gamma$ and $B_4$ for different phase transitions \cite{Pelissetto:2000ek}.}
    \label{table:critical_exponents_values}
  \end{center}
\end{table}
A true non-analytic phase transition can only exist in the infinite volume limit, therefore to extract this transition from simulations of finite volumes, an extrapolation with a finite size scaling is needed. One way is to use the kurtosis $B_4$ for approaching the infinite volume limit. The critical value of $B_4$ in the thermodynamic limit for different orders of the phase transition is given in table \ref{table:critical_exponents_values}. The leading finite size corrections are obtained by performing a Taylor expansion 
about a critical point in infinite volume, to which we fit our data,
\begin{align}
B_4(\kappa,N_s)=B_4(\kappa_c,\infty)+a_1(\kappa-\kappa_c)N_s^{1/\nu}+\cdots .
\label{eq:kurtosis_fit}
\end{align}

\begin{figure}[t]
\begin{center}
	\begin{subfigure}{.5\textwidth}
\centering
	\scalebox{.8}{
\begingroup
  \makeatletter
  \providecommand\color[2][]{%
    \GenericError{(gnuplot) \space\space\space\@spaces}{%
      Package color not loaded in conjunction with
      terminal option `colourtext'%
    }{See the gnuplot documentation for explanation.%
    }{Either use 'blacktext' in gnuplot or load the package
      color.sty in LaTeX.}%
    \renewcommand\color[2][]{}%
  }%
  \providecommand\includegraphics[2][]{%
    \GenericError{(gnuplot) \space\space\space\@spaces}{%
      Package graphicx or graphics not loaded%
    }{See the gnuplot documentation for explanation.%
    }{The gnuplot epslatex terminal needs graphicx.sty or graphics.sty.}%
    \renewcommand\includegraphics[2][]{}%
  }%
  \providecommand\rotatebox[2]{#2}%
  \@ifundefined{ifGPcolor}{%
    \newif\ifGPcolor
    \GPcolortrue
  }{}%
  \@ifundefined{ifGPblacktext}{%
    \newif\ifGPblacktext
    \GPblacktexttrue
  }{}%
  \let\gplgaddtomacro\g@addto@macro
  \gdef\gplbacktext{}%
  \gdef\gplfronttext{}%
  \makeatother
  \ifGPblacktext
    \def\colorrgb#1{}%
    \def\colorgray#1{}%
  \else
    \ifGPcolor
      \def\colorrgb#1{\color[rgb]{#1}}%
      \def\colorgray#1{\color[gray]{#1}}%
      \expandafter\def\csname LTw\endcsname{\color{white}}%
      \expandafter\def\csname LTb\endcsname{\color{black}}%
      \expandafter\def\csname LTa\endcsname{\color{black}}%
      \expandafter\def\csname LT0\endcsname{\color[rgb]{1,0,0}}%
      \expandafter\def\csname LT1\endcsname{\color[rgb]{0,1,0}}%
      \expandafter\def\csname LT2\endcsname{\color[rgb]{0,0,1}}%
      \expandafter\def\csname LT3\endcsname{\color[rgb]{1,0,1}}%
      \expandafter\def\csname LT4\endcsname{\color[rgb]{0,1,1}}%
      \expandafter\def\csname LT5\endcsname{\color[rgb]{1,1,0}}%
      \expandafter\def\csname LT6\endcsname{\color[rgb]{0,0,0}}%
      \expandafter\def\csname LT7\endcsname{\color[rgb]{1,0.3,0}}%
      \expandafter\def\csname LT8\endcsname{\color[rgb]{0.5,0.5,0.5}}%
    \else
      \def\colorrgb#1{\color{black}}%
      \def\colorgray#1{\color[gray]{#1}}%
      \expandafter\def\csname LTw\endcsname{\color{white}}%
      \expandafter\def\csname LTb\endcsname{\color{black}}%
      \expandafter\def\csname LTa\endcsname{\color{black}}%
      \expandafter\def\csname LT0\endcsname{\color{black}}%
      \expandafter\def\csname LT1\endcsname{\color{black}}%
      \expandafter\def\csname LT2\endcsname{\color{black}}%
      \expandafter\def\csname LT3\endcsname{\color{black}}%
      \expandafter\def\csname LT4\endcsname{\color{black}}%
      \expandafter\def\csname LT5\endcsname{\color{black}}%
      \expandafter\def\csname LT6\endcsname{\color{black}}%
      \expandafter\def\csname LT7\endcsname{\color{black}}%
      \expandafter\def\csname LT8\endcsname{\color{black}}%
    \fi
  \fi
    \setlength{\unitlength}{0.0500bp}%
    \ifx\gptboxheight\undefined%
      \newlength{\gptboxheight}%
      \newlength{\gptboxwidth}%
      \newsavebox{\gptboxtext}%
    \fi%
    \setlength{\fboxrule}{0.5pt}%
    \setlength{\fboxsep}{1pt}%
\begin{picture}(5102.00,3400.00)%
    \gplgaddtomacro\gplbacktext{%
      \csname LTb\endcsname
      \put(1210,995){\makebox(0,0)[r]{\strut{}$0.4286$}}%
      \put(1210,1723){\makebox(0,0)[r]{\strut{}$0.4287$}}%
      \put(1210,2451){\makebox(0,0)[r]{\strut{}$0.4288$}}%
      \put(1210,3179){\makebox(0,0)[r]{\strut{}$0.4289$}}%
      \put(1903,484){\makebox(0,0){\strut{}$0.06$}}%
      \put(3024,484){\makebox(0,0){\strut{}$0.07$}}%
      \put(4145,484){\makebox(0,0){\strut{}$0.08$}}%
    }%
    \gplgaddtomacro\gplfronttext{%
      \csname LTb\endcsname
      \put(209,1941){\rotatebox{-270}{\makebox(0,0){\strut{}$u(\beta)$}}}%
      \put(3023,154){\makebox(0,0){\strut{}$\kappa$}}%
      \csname LTb\endcsname
      \put(3470,2945){\makebox(0,0)[l]{\strut{}$u_{\mathrm{pc}}(\kappa)$}}%
      \csname LTb\endcsname
      \put(3470,2725){\makebox(0,0)[l]{\strut{}$\mathrm{linear\, fit}$}}%
      \csname LTb\endcsname
      \put(3470,2505){\makebox(0,0)[l]{\strut{}$\mathrm{critical\, point}$}}%
    }%
    \gplbacktext
    \put(0,0){\includegraphics[width={255.10bp},height={170.00bp}]{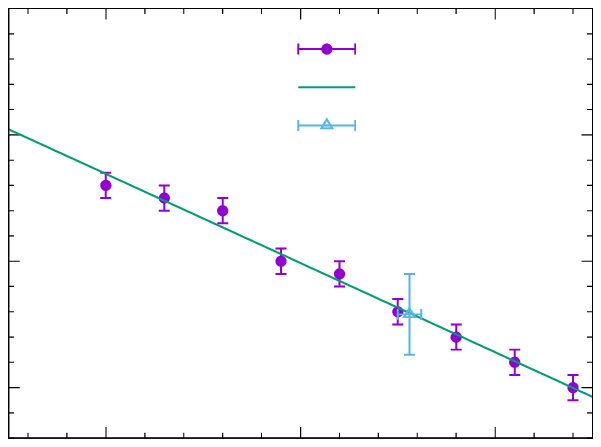}}%
    \gplfronttext
  \end{picture}%
\endgroup
}
	\caption{$S_{\text{eff}}^g+S_0+S_2, N_f=1, N_t=4$}
	\label{fig:pseudoline_Nf1Nt4}
\end{subfigure}%
\begin{subfigure}{.5\textwidth}
\centering
	\scalebox{.8}{
\begingroup
  \makeatletter
  \providecommand\color[2][]{%
    \GenericError{(gnuplot) \space\space\space\@spaces}{%
      Package color not loaded in conjunction with
      terminal option `colourtext'%
    }{See the gnuplot documentation for explanation.%
    }{Either use 'blacktext' in gnuplot or load the package
      color.sty in LaTeX.}%
    \renewcommand\color[2][]{}%
  }%
  \providecommand\includegraphics[2][]{%
    \GenericError{(gnuplot) \space\space\space\@spaces}{%
      Package graphicx or graphics not loaded%
    }{See the gnuplot documentation for explanation.%
    }{The gnuplot epslatex terminal needs graphicx.sty or graphics.sty.}%
    \renewcommand\includegraphics[2][]{}%
  }%
  \providecommand\rotatebox[2]{#2}%
  \@ifundefined{ifGPcolor}{%
    \newif\ifGPcolor
    \GPcolortrue
  }{}%
  \@ifundefined{ifGPblacktext}{%
    \newif\ifGPblacktext
    \GPblacktexttrue
  }{}%
  \let\gplgaddtomacro\g@addto@macro
  \gdef\gplbacktext{}%
  \gdef\gplfronttext{}%
  \makeatother
  \ifGPblacktext
    \def\colorrgb#1{}%
    \def\colorgray#1{}%
  \else
    \ifGPcolor
      \def\colorrgb#1{\color[rgb]{#1}}%
      \def\colorgray#1{\color[gray]{#1}}%
      \expandafter\def\csname LTw\endcsname{\color{white}}%
      \expandafter\def\csname LTb\endcsname{\color{black}}%
      \expandafter\def\csname LTa\endcsname{\color{black}}%
      \expandafter\def\csname LT0\endcsname{\color[rgb]{1,0,0}}%
      \expandafter\def\csname LT1\endcsname{\color[rgb]{0,1,0}}%
      \expandafter\def\csname LT2\endcsname{\color[rgb]{0,0,1}}%
      \expandafter\def\csname LT3\endcsname{\color[rgb]{1,0,1}}%
      \expandafter\def\csname LT4\endcsname{\color[rgb]{0,1,1}}%
      \expandafter\def\csname LT5\endcsname{\color[rgb]{1,1,0}}%
      \expandafter\def\csname LT6\endcsname{\color[rgb]{0,0,0}}%
      \expandafter\def\csname LT7\endcsname{\color[rgb]{1,0.3,0}}%
      \expandafter\def\csname LT8\endcsname{\color[rgb]{0.5,0.5,0.5}}%
    \else
      \def\colorrgb#1{\color{black}}%
      \def\colorgray#1{\color[gray]{#1}}%
      \expandafter\def\csname LTw\endcsname{\color{white}}%
      \expandafter\def\csname LTb\endcsname{\color{black}}%
      \expandafter\def\csname LTa\endcsname{\color{black}}%
      \expandafter\def\csname LT0\endcsname{\color{black}}%
      \expandafter\def\csname LT1\endcsname{\color{black}}%
      \expandafter\def\csname LT2\endcsname{\color{black}}%
      \expandafter\def\csname LT3\endcsname{\color{black}}%
      \expandafter\def\csname LT4\endcsname{\color{black}}%
      \expandafter\def\csname LT5\endcsname{\color{black}}%
      \expandafter\def\csname LT6\endcsname{\color{black}}%
      \expandafter\def\csname LT7\endcsname{\color{black}}%
      \expandafter\def\csname LT8\endcsname{\color{black}}%
    \fi
  \fi
    \setlength{\unitlength}{0.0500bp}%
    \ifx\gptboxheight\undefined%
      \newlength{\gptboxheight}%
      \newlength{\gptboxwidth}%
      \newsavebox{\gptboxtext}%
    \fi%
    \setlength{\fboxrule}{0.5pt}%
    \setlength{\fboxsep}{1pt}%
\begin{picture}(5102.00,3400.00)%
    \gplgaddtomacro\gplbacktext{%
      \csname LTb\endcsname
      \put(1210,1034){\makebox(0,0)[r]{\strut{}$0.4286$}}%
      \put(1210,1859){\makebox(0,0)[r]{\strut{}$0.4287$}}%
      \put(1210,2684){\makebox(0,0)[r]{\strut{}$0.4288$}}%
      \put(1342,484){\makebox(0,0){\strut{}$0.05$}}%
      \put(2393,484){\makebox(0,0){\strut{}$0.06$}}%
      \put(3444,484){\makebox(0,0){\strut{}$0.07$}}%
      \put(4495,484){\makebox(0,0){\strut{}$0.08$}}%
    }%
    \gplgaddtomacro\gplfronttext{%
      \csname LTb\endcsname
      \put(209,1941){\rotatebox{-270}{\makebox(0,0){\strut{}$u(\beta)$}}}%
      \put(3023,154){\makebox(0,0){\strut{}$\kappa$}}%
      \csname LTb\endcsname
      \put(2125,1460){\makebox(0,0)[l]{\strut{}$u_{\mathrm{pc}}(\kappa)$}}%
      \csname LTb\endcsname
      \put(2125,1240){\makebox(0,0)[l]{\strut{}$\mathrm{linear\, fit}$}}%
      \csname LTb\endcsname
      \put(2125,1020){\makebox(0,0)[l]{\strut{}$\mathrm{critical\, point}$}}%
    }%
    \gplbacktext
    \put(0,0){\includegraphics[width={255.10bp},height={170.00bp}]{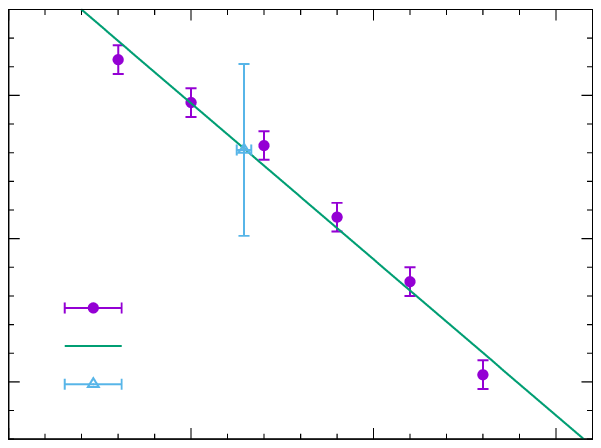}}%
    \gplfronttext
  \end{picture}%
\endgroup
}
	\caption{$S_{\text{eff}}^g+S_0+S_2, N_f=2, N_t=4$}
	\label{fig:pseudoline_Nf2Nt4}
\end{subfigure}
\begin{subfigure}{.5\textwidth}
\centering
	\scalebox{.8}{
\begingroup
  \makeatletter
  \providecommand\color[2][]{%
    \GenericError{(gnuplot) \space\space\space\@spaces}{%
      Package color not loaded in conjunction with
      terminal option `colourtext'%
    }{See the gnuplot documentation for explanation.%
    }{Either use 'blacktext' in gnuplot or load the package
      color.sty in LaTeX.}%
    \renewcommand\color[2][]{}%
  }%
  \providecommand\includegraphics[2][]{%
    \GenericError{(gnuplot) \space\space\space\@spaces}{%
      Package graphicx or graphics not loaded%
    }{See the gnuplot documentation for explanation.%
    }{The gnuplot epslatex terminal needs graphicx.sty or graphics.sty.}%
    \renewcommand\includegraphics[2][]{}%
  }%
  \providecommand\rotatebox[2]{#2}%
  \@ifundefined{ifGPcolor}{%
    \newif\ifGPcolor
    \GPcolortrue
  }{}%
  \@ifundefined{ifGPblacktext}{%
    \newif\ifGPblacktext
    \GPblacktexttrue
  }{}%
  \let\gplgaddtomacro\g@addto@macro
  \gdef\gplbacktext{}%
  \gdef\gplfronttext{}%
  \makeatother
  \ifGPblacktext
    \def\colorrgb#1{}%
    \def\colorgray#1{}%
  \else
    \ifGPcolor
      \def\colorrgb#1{\color[rgb]{#1}}%
      \def\colorgray#1{\color[gray]{#1}}%
      \expandafter\def\csname LTw\endcsname{\color{white}}%
      \expandafter\def\csname LTb\endcsname{\color{black}}%
      \expandafter\def\csname LTa\endcsname{\color{black}}%
      \expandafter\def\csname LT0\endcsname{\color[rgb]{1,0,0}}%
      \expandafter\def\csname LT1\endcsname{\color[rgb]{0,1,0}}%
      \expandafter\def\csname LT2\endcsname{\color[rgb]{0,0,1}}%
      \expandafter\def\csname LT3\endcsname{\color[rgb]{1,0,1}}%
      \expandafter\def\csname LT4\endcsname{\color[rgb]{0,1,1}}%
      \expandafter\def\csname LT5\endcsname{\color[rgb]{1,1,0}}%
      \expandafter\def\csname LT6\endcsname{\color[rgb]{0,0,0}}%
      \expandafter\def\csname LT7\endcsname{\color[rgb]{1,0.3,0}}%
      \expandafter\def\csname LT8\endcsname{\color[rgb]{0.5,0.5,0.5}}%
    \else
      \def\colorrgb#1{\color{black}}%
      \def\colorgray#1{\color[gray]{#1}}%
      \expandafter\def\csname LTw\endcsname{\color{white}}%
      \expandafter\def\csname LTb\endcsname{\color{black}}%
      \expandafter\def\csname LTa\endcsname{\color{black}}%
      \expandafter\def\csname LT0\endcsname{\color{black}}%
      \expandafter\def\csname LT1\endcsname{\color{black}}%
      \expandafter\def\csname LT2\endcsname{\color{black}}%
      \expandafter\def\csname LT3\endcsname{\color{black}}%
      \expandafter\def\csname LT4\endcsname{\color{black}}%
      \expandafter\def\csname LT5\endcsname{\color{black}}%
      \expandafter\def\csname LT6\endcsname{\color{black}}%
      \expandafter\def\csname LT7\endcsname{\color{black}}%
      \expandafter\def\csname LT8\endcsname{\color{black}}%
    \fi
  \fi
    \setlength{\unitlength}{0.0500bp}%
    \ifx\gptboxheight\undefined%
      \newlength{\gptboxheight}%
      \newlength{\gptboxwidth}%
      \newsavebox{\gptboxtext}%
    \fi%
    \setlength{\fboxrule}{0.5pt}%
    \setlength{\fboxsep}{1pt}%
\begin{picture}(5102.00,3400.00)%
    \gplgaddtomacro\gplbacktext{%
      \csname LTb\endcsname
      \put(1342,704){\makebox(0,0)[r]{\strut{}$0.4431$}}%
      \put(1342,1323){\makebox(0,0)[r]{\strut{}$0.44315$}}%
      \put(1342,1941){\makebox(0,0)[r]{\strut{}$0.4432$}}%
      \put(1342,2560){\makebox(0,0)[r]{\strut{}$0.44325$}}%
      \put(1342,3179){\makebox(0,0)[r]{\strut{}$0.4433$}}%
      \put(1854,484){\makebox(0,0){\strut{}$0$}}%
      \put(2614,484){\makebox(0,0){\strut{}$0.04$}}%
      \put(3375,484){\makebox(0,0){\strut{}$0.08$}}%
      \put(4135,484){\makebox(0,0){\strut{}$0.12$}}%
    }%
    \gplgaddtomacro\gplfronttext{%
      \csname LTb\endcsname
      \put(209,1941){\rotatebox{-270}{\makebox(0,0){\strut{}$u(\beta)$}}}%
      \put(3089,154){\makebox(0,0){\strut{}$\kappa$}}%
      \csname LTb\endcsname
      \put(2885,1584){\makebox(0,0)[l]{\strut{}$u(\kappa)$}}%
    }%
    \gplbacktext
    \put(0,0){\includegraphics[width={255.10bp},height={170.00bp}]{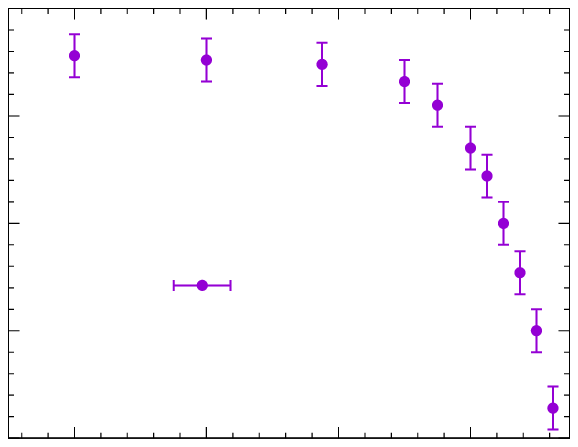}}%
    \gplfronttext
  \end{picture}%
\endgroup
}
	\caption{$S_{\text{eff}}^g+S_0+S_2, N_f=1, N_t=6$}
	\label{fig:pseudoline_Nf1Nt6}
\end{subfigure}%
\begin{subfigure}{.5\textwidth}
\centering
	\scalebox{.8}{
\begingroup
  \makeatletter
  \providecommand\color[2][]{%
    \GenericError{(gnuplot) \space\space\space\@spaces}{%
      Package color not loaded in conjunction with
      terminal option `colourtext'%
    }{See the gnuplot documentation for explanation.%
    }{Either use 'blacktext' in gnuplot or load the package
      color.sty in LaTeX.}%
    \renewcommand\color[2][]{}%
  }%
  \providecommand\includegraphics[2][]{%
    \GenericError{(gnuplot) \space\space\space\@spaces}{%
      Package graphicx or graphics not loaded%
    }{See the gnuplot documentation for explanation.%
    }{The gnuplot epslatex terminal needs graphicx.sty or graphics.sty.}%
    \renewcommand\includegraphics[2][]{}%
  }%
  \providecommand\rotatebox[2]{#2}%
  \@ifundefined{ifGPcolor}{%
    \newif\ifGPcolor
    \GPcolortrue
  }{}%
  \@ifundefined{ifGPblacktext}{%
    \newif\ifGPblacktext
    \GPblacktexttrue
  }{}%
  \let\gplgaddtomacro\g@addto@macro
  \gdef\gplbacktext{}%
  \gdef\gplfronttext{}%
  \makeatother
  \ifGPblacktext
    \def\colorrgb#1{}%
    \def\colorgray#1{}%
  \else
    \ifGPcolor
      \def\colorrgb#1{\color[rgb]{#1}}%
      \def\colorgray#1{\color[gray]{#1}}%
      \expandafter\def\csname LTw\endcsname{\color{white}}%
      \expandafter\def\csname LTb\endcsname{\color{black}}%
      \expandafter\def\csname LTa\endcsname{\color{black}}%
      \expandafter\def\csname LT0\endcsname{\color[rgb]{1,0,0}}%
      \expandafter\def\csname LT1\endcsname{\color[rgb]{0,1,0}}%
      \expandafter\def\csname LT2\endcsname{\color[rgb]{0,0,1}}%
      \expandafter\def\csname LT3\endcsname{\color[rgb]{1,0,1}}%
      \expandafter\def\csname LT4\endcsname{\color[rgb]{0,1,1}}%
      \expandafter\def\csname LT5\endcsname{\color[rgb]{1,1,0}}%
      \expandafter\def\csname LT6\endcsname{\color[rgb]{0,0,0}}%
      \expandafter\def\csname LT7\endcsname{\color[rgb]{1,0.3,0}}%
      \expandafter\def\csname LT8\endcsname{\color[rgb]{0.5,0.5,0.5}}%
    \else
      \def\colorrgb#1{\color{black}}%
      \def\colorgray#1{\color[gray]{#1}}%
      \expandafter\def\csname LTw\endcsname{\color{white}}%
      \expandafter\def\csname LTb\endcsname{\color{black}}%
      \expandafter\def\csname LTa\endcsname{\color{black}}%
      \expandafter\def\csname LT0\endcsname{\color{black}}%
      \expandafter\def\csname LT1\endcsname{\color{black}}%
      \expandafter\def\csname LT2\endcsname{\color{black}}%
      \expandafter\def\csname LT3\endcsname{\color{black}}%
      \expandafter\def\csname LT4\endcsname{\color{black}}%
      \expandafter\def\csname LT5\endcsname{\color{black}}%
      \expandafter\def\csname LT6\endcsname{\color{black}}%
      \expandafter\def\csname LT7\endcsname{\color{black}}%
      \expandafter\def\csname LT8\endcsname{\color{black}}%
    \fi
  \fi
    \setlength{\unitlength}{0.0500bp}%
    \ifx\gptboxheight\undefined%
      \newlength{\gptboxheight}%
      \newlength{\gptboxwidth}%
      \newsavebox{\gptboxtext}%
    \fi%
    \setlength{\fboxrule}{0.5pt}%
    \setlength{\fboxsep}{1pt}%
\begin{picture}(5102.00,3400.00)%
    \gplgaddtomacro\gplbacktext{%
      \csname LTb\endcsname
      \put(1210,704){\makebox(0,0)[r]{\strut{}$0.443$}}%
      \put(1210,1529){\makebox(0,0)[r]{\strut{}$0.4431$}}%
      \put(1210,2354){\makebox(0,0)[r]{\strut{}$0.4432$}}%
      \put(1210,3179){\makebox(0,0)[r]{\strut{}$0.4433$}}%
      \put(1566,484){\makebox(0,0){\strut{}$0$}}%
      \put(2463,484){\makebox(0,0){\strut{}$0.04$}}%
      \put(3360,484){\makebox(0,0){\strut{}$0.08$}}%
      \put(4257,484){\makebox(0,0){\strut{}$0.12$}}%
    }%
    \gplgaddtomacro\gplfronttext{%
      \csname LTb\endcsname
      \put(209,1941){\rotatebox{-270}{\makebox(0,0){\strut{}$u(\beta)$}}}%
      \put(3023,154){\makebox(0,0){\strut{}$\kappa$}}%
      \csname LTb\endcsname
      \put(2832,1584){\makebox(0,0)[l]{\strut{}$u(\kappa)$}}%
    }%
    \gplbacktext
    \put(0,0){\includegraphics[width={255.10bp},height={170.00bp}]{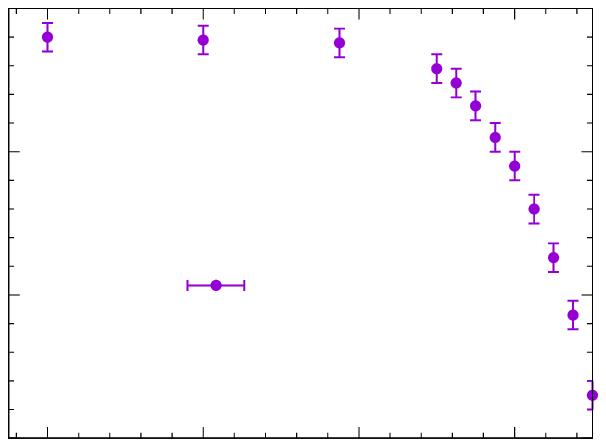}}%
    \gplfronttext
  \end{picture}%
\endgroup
}
	\caption{$S_{\text{eff}}^g+S_0+S_2, N_f=2, N_t=6$}
	\label{fig:pseudoline_Nf2Nt6}
\end{subfigure}
\caption{The pseudo-critical line for the next-to-leading order effective theory with $N_f=1,2$, $N_t=4,6$ and $N_s=24$. A linear fit was performed for $N_t=4$ according to equation \eqref{eq:pseudoline_fit}.} 
\label{fig:u_kappa_Nt4_linearfit}
\end{center}
\end {figure}

Our investigation proceeds in two steps: first, the pseudo-critical line $u_{pc}(\kappa)$ is mapped, subsequently its critical point $(u_c,\kappa_c)$ is located. The pseudo-critical line is found by fixing values of $\kappa$ and performing a 
$u$-scan at $N_s=16,20,24$, identifying the maximum of the susceptibility and minimum $B_4$. Results on $N_s=24$ are shown for $N_f=1$ in figure \ref{fig:pseudoline_Nf1Nt4} and for $N_f=2$ in
 \ref{fig:pseudoline_Nf2Nt4}. Due to the smallness of $\kappa$, the entire line can be parametrised as  \cite{Alford:2001ug}
\begin{align}
u_{pc}(\kappa)=u_0-a_1\kappa,
\label{eq:pseudoline_fit}
\end{align}
with the following fit results,
\begin{align}
a_1&=0.0071(3), \hspace{0.5cm} u_0=0.429192(22) , \hspace{0.5cm} \chi^2/\text{d.o.f}=0.49,\hspace{0.5cm} \text{for} \hspace{0.5cm} N_f=1,\\
a_1&=0.0109(8), \hspace{0.5cm} u_0=0.429448(52) , \hspace{0.5cm} \chi^2/\text{d.o.f}=1.74,\hspace{0.5cm} \text{for} \hspace{0.5cm} N_f=2.
\end{align}
However, the linearity does not hold for $N_t=6$ and $N_f=1,2$ as illustrated in figure \ref{fig:pseudoline_Nf1Nt6} and \ref{fig:pseudoline_Nf2Nt6}.

\begin {figure}[t]
\label{table:B4fitting}
\begin{center}
	\begin{subfigure}{.5\textwidth}
\centering
	\scalebox{.9}{
\begingroup
  \makeatletter
  \providecommand\color[2][]{%
    \GenericError{(gnuplot) \space\space\space\@spaces}{%
      Package color not loaded in conjunction with
      terminal option `colourtext'%
    }{See the gnuplot documentation for explanation.%
    }{Either use 'blacktext' in gnuplot or load the package
      color.sty in LaTeX.}%
    \renewcommand\color[2][]{}%
  }%
  \providecommand\includegraphics[2][]{%
    \GenericError{(gnuplot) \space\space\space\@spaces}{%
      Package graphicx or graphics not loaded%
    }{See the gnuplot documentation for explanation.%
    }{The gnuplot epslatex terminal needs graphicx.sty or graphics.sty.}%
    \renewcommand\includegraphics[2][]{}%
  }%
  \providecommand\rotatebox[2]{#2}%
  \@ifundefined{ifGPcolor}{%
    \newif\ifGPcolor
    \GPcolortrue
  }{}%
  \@ifundefined{ifGPblacktext}{%
    \newif\ifGPblacktext
    \GPblacktexttrue
  }{}%
  \let\gplgaddtomacro\g@addto@macro
  \gdef\gplbacktext{}%
  \gdef\gplfronttext{}%
  \makeatother
  \ifGPblacktext
    \def\colorrgb#1{}%
    \def\colorgray#1{}%
  \else
    \ifGPcolor
      \def\colorrgb#1{\color[rgb]{#1}}%
      \def\colorgray#1{\color[gray]{#1}}%
      \expandafter\def\csname LTw\endcsname{\color{white}}%
      \expandafter\def\csname LTb\endcsname{\color{black}}%
      \expandafter\def\csname LTa\endcsname{\color{black}}%
      \expandafter\def\csname LT0\endcsname{\color[rgb]{1,0,0}}%
      \expandafter\def\csname LT1\endcsname{\color[rgb]{0,1,0}}%
      \expandafter\def\csname LT2\endcsname{\color[rgb]{0,0,1}}%
      \expandafter\def\csname LT3\endcsname{\color[rgb]{1,0,1}}%
      \expandafter\def\csname LT4\endcsname{\color[rgb]{0,1,1}}%
      \expandafter\def\csname LT5\endcsname{\color[rgb]{1,1,0}}%
      \expandafter\def\csname LT6\endcsname{\color[rgb]{0,0,0}}%
      \expandafter\def\csname LT7\endcsname{\color[rgb]{1,0.3,0}}%
      \expandafter\def\csname LT8\endcsname{\color[rgb]{0.5,0.5,0.5}}%
    \else
      \def\colorrgb#1{\color{black}}%
      \def\colorgray#1{\color[gray]{#1}}%
      \expandafter\def\csname LTw\endcsname{\color{white}}%
      \expandafter\def\csname LTb\endcsname{\color{black}}%
      \expandafter\def\csname LTa\endcsname{\color{black}}%
      \expandafter\def\csname LT0\endcsname{\color{black}}%
      \expandafter\def\csname LT1\endcsname{\color{black}}%
      \expandafter\def\csname LT2\endcsname{\color{black}}%
      \expandafter\def\csname LT3\endcsname{\color{black}}%
      \expandafter\def\csname LT4\endcsname{\color{black}}%
      \expandafter\def\csname LT5\endcsname{\color{black}}%
      \expandafter\def\csname LT6\endcsname{\color{black}}%
      \expandafter\def\csname LT7\endcsname{\color{black}}%
      \expandafter\def\csname LT8\endcsname{\color{black}}%
    \fi
  \fi
    \setlength{\unitlength}{0.0500bp}%
    \ifx\gptboxheight\undefined%
      \newlength{\gptboxheight}%
      \newlength{\gptboxwidth}%
      \newsavebox{\gptboxtext}%
    \fi%
    \setlength{\fboxrule}{0.5pt}%
    \setlength{\fboxsep}{1pt}%
\begin{picture}(5102.00,3400.00)%
    \gplgaddtomacro\gplbacktext{%
      \csname LTb\endcsname
      \put(814,704){\makebox(0,0)[r]{\strut{}$1$}}%
      \put(814,1529){\makebox(0,0)[r]{\strut{}$1.5$}}%
      \put(814,2354){\makebox(0,0)[r]{\strut{}$2$}}%
      \put(814,3179){\makebox(0,0)[r]{\strut{}$2.5$}}%
      \put(1930,484){\makebox(0,0){\strut{}$0.0006$}}%
      \put(3273,484){\makebox(0,0){\strut{}$0.0009$}}%
      \put(4616,484){\makebox(0,0){\strut{}$0.0012$}}%
      \put(1259,2849){\makebox(0,0)[l]{\strut{}$\chi^2/{\mathrm{d.o.f}}=1.49$}}%
    }%
    \gplgaddtomacro\gplfronttext{%
      \csname LTb\endcsname
      \put(209,1941){\rotatebox{-270}{\makebox(0,0){\strut{}$B_4(\lambda_c,h_1, N_s)$}}}%
      \put(2825,154){\makebox(0,0){\strut{}$h_{1c}$}}%
      \csname LTb\endcsname
      \put(3604,1337){\makebox(0,0)[l]{\strut{}16x16x16}}%
      \csname LTb\endcsname
      \put(3604,1117){\makebox(0,0)[l]{\strut{}24x24x24}}%
      \csname LTb\endcsname
      \put(3604,897){\makebox(0,0)[l]{\strut{}32x32x32}}%
    }%
    \gplbacktext
    \put(0,0){\includegraphics[width={255.10bp},height={170.00bp}]{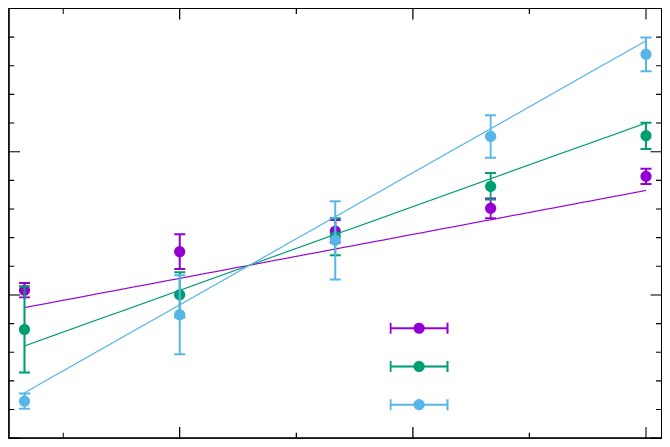}}%
    \gplfronttext
  \end{picture}%
\endgroup
}
	\caption{$S_{\text{eff}}^g+S_0$, $N_f=1, N_t=4$, ($(\lambda_c,h_{1c})$ can be converted to $(u_c,\kappa_c)$)}
	\label{fig:B4fit_k1Nf1Nt4}
\end{subfigure}%
\begin{subfigure}{.5\textwidth}
\centering
	\scalebox{.9}{
\begingroup
  \makeatletter
  \providecommand\color[2][]{%
    \GenericError{(gnuplot) \space\space\space\@spaces}{%
      Package color not loaded in conjunction with
      terminal option `colourtext'%
    }{See the gnuplot documentation for explanation.%
    }{Either use 'blacktext' in gnuplot or load the package
      color.sty in LaTeX.}%
    \renewcommand\color[2][]{}%
  }%
  \providecommand\includegraphics[2][]{%
    \GenericError{(gnuplot) \space\space\space\@spaces}{%
      Package graphicx or graphics not loaded%
    }{See the gnuplot documentation for explanation.%
    }{The gnuplot epslatex terminal needs graphicx.sty or graphics.sty.}%
    \renewcommand\includegraphics[2][]{}%
  }%
  \providecommand\rotatebox[2]{#2}%
  \@ifundefined{ifGPcolor}{%
    \newif\ifGPcolor
    \GPcolortrue
  }{}%
  \@ifundefined{ifGPblacktext}{%
    \newif\ifGPblacktext
    \GPblacktexttrue
  }{}%
  \let\gplgaddtomacro\g@addto@macro
  \gdef\gplbacktext{}%
  \gdef\gplfronttext{}%
  \makeatother
  \ifGPblacktext
    \def\colorrgb#1{}%
    \def\colorgray#1{}%
  \else
    \ifGPcolor
      \def\colorrgb#1{\color[rgb]{#1}}%
      \def\colorgray#1{\color[gray]{#1}}%
      \expandafter\def\csname LTw\endcsname{\color{white}}%
      \expandafter\def\csname LTb\endcsname{\color{black}}%
      \expandafter\def\csname LTa\endcsname{\color{black}}%
      \expandafter\def\csname LT0\endcsname{\color[rgb]{1,0,0}}%
      \expandafter\def\csname LT1\endcsname{\color[rgb]{0,1,0}}%
      \expandafter\def\csname LT2\endcsname{\color[rgb]{0,0,1}}%
      \expandafter\def\csname LT3\endcsname{\color[rgb]{1,0,1}}%
      \expandafter\def\csname LT4\endcsname{\color[rgb]{0,1,1}}%
      \expandafter\def\csname LT5\endcsname{\color[rgb]{1,1,0}}%
      \expandafter\def\csname LT6\endcsname{\color[rgb]{0,0,0}}%
      \expandafter\def\csname LT7\endcsname{\color[rgb]{1,0.3,0}}%
      \expandafter\def\csname LT8\endcsname{\color[rgb]{0.5,0.5,0.5}}%
    \else
      \def\colorrgb#1{\color{black}}%
      \def\colorgray#1{\color[gray]{#1}}%
      \expandafter\def\csname LTw\endcsname{\color{white}}%
      \expandafter\def\csname LTb\endcsname{\color{black}}%
      \expandafter\def\csname LTa\endcsname{\color{black}}%
      \expandafter\def\csname LT0\endcsname{\color{black}}%
      \expandafter\def\csname LT1\endcsname{\color{black}}%
      \expandafter\def\csname LT2\endcsname{\color{black}}%
      \expandafter\def\csname LT3\endcsname{\color{black}}%
      \expandafter\def\csname LT4\endcsname{\color{black}}%
      \expandafter\def\csname LT5\endcsname{\color{black}}%
      \expandafter\def\csname LT6\endcsname{\color{black}}%
      \expandafter\def\csname LT7\endcsname{\color{black}}%
      \expandafter\def\csname LT8\endcsname{\color{black}}%
    \fi
  \fi
    \setlength{\unitlength}{0.0500bp}%
    \ifx\gptboxheight\undefined%
      \newlength{\gptboxheight}%
      \newlength{\gptboxwidth}%
      \newsavebox{\gptboxtext}%
    \fi%
    \setlength{\fboxrule}{0.5pt}%
    \setlength{\fboxsep}{1pt}%
\begin{picture}(5102.00,3400.00)%
    \gplgaddtomacro\gplbacktext{%
      \csname LTb\endcsname
      \put(814,704){\makebox(0,0)[r]{\strut{}$1.2$}}%
      \put(814,1529){\makebox(0,0)[r]{\strut{}$1.5$}}%
      \put(814,2354){\makebox(0,0)[r]{\strut{}$1.8$}}%
      \put(814,3179){\makebox(0,0)[r]{\strut{}$2.1$}}%
      \put(1117,484){\makebox(0,0){\strut{}$0.125$}}%
      \put(1971,484){\makebox(0,0){\strut{}$0.13$}}%
      \put(2826,484){\makebox(0,0){\strut{}$0.135$}}%
      \put(3680,484){\makebox(0,0){\strut{}$0.14$}}%
      \put(4534,484){\makebox(0,0){\strut{}$0.145$}}%
      \put(1117,2904){\makebox(0,0)[l]{\strut{}$\chi^2/{\mathrm{d.o.f}}=0.99$}}%
    }%
    \gplgaddtomacro\gplfronttext{%
      \csname LTb\endcsname
      \put(209,1941){\rotatebox{-270}{\makebox(0,0){\strut{}$B_4(u_c,\kappa, N_s)$}}}%
      \put(2825,154){\makebox(0,0){\strut{}$\kappa_c$}}%
      \csname LTb\endcsname
      \put(3604,1337){\makebox(0,0)[l]{\strut{}16x16x16}}%
      \csname LTb\endcsname
      \put(3604,1117){\makebox(0,0)[l]{\strut{}20x20x20}}%
      \csname LTb\endcsname
      \put(3604,897){\makebox(0,0)[l]{\strut{}24x24x24}}%
    }%
    \gplbacktext
    \put(0,0){\includegraphics[width={255.10bp},height={170.00bp}]{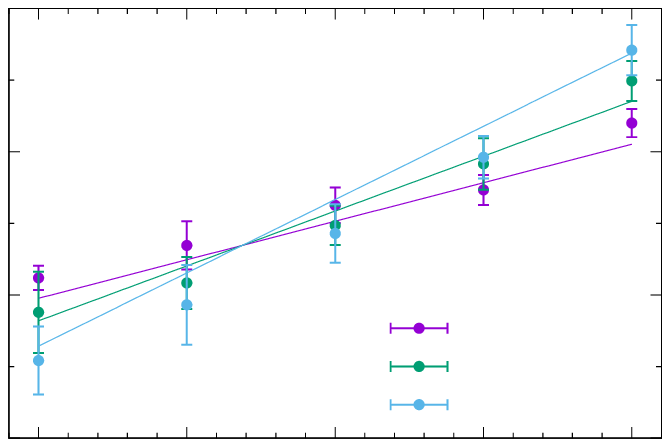}}%
    \gplfronttext
  \end{picture}%
\endgroup
}
	\caption{$S_{\text{eff}}^g+S_0+S_2, N_f=1, N_t=6$}
\end{subfigure}
\begin{subfigure}{.5\textwidth}
\centering
	\scalebox{.9}{
\begingroup
  \makeatletter
  \providecommand\color[2][]{%
    \GenericError{(gnuplot) \space\space\space\@spaces}{%
      Package color not loaded in conjunction with
      terminal option `colourtext'%
    }{See the gnuplot documentation for explanation.%
    }{Either use 'blacktext' in gnuplot or load the package
      color.sty in LaTeX.}%
    \renewcommand\color[2][]{}%
  }%
  \providecommand\includegraphics[2][]{%
    \GenericError{(gnuplot) \space\space\space\@spaces}{%
      Package graphicx or graphics not loaded%
    }{See the gnuplot documentation for explanation.%
    }{The gnuplot epslatex terminal needs graphicx.sty or graphics.sty.}%
    \renewcommand\includegraphics[2][]{}%
  }%
  \providecommand\rotatebox[2]{#2}%
  \@ifundefined{ifGPcolor}{%
    \newif\ifGPcolor
    \GPcolortrue
  }{}%
  \@ifundefined{ifGPblacktext}{%
    \newif\ifGPblacktext
    \GPblacktexttrue
  }{}%
  \let\gplgaddtomacro\g@addto@macro
  \gdef\gplbacktext{}%
  \gdef\gplfronttext{}%
  \makeatother
  \ifGPblacktext
    \def\colorrgb#1{}%
    \def\colorgray#1{}%
  \else
    \ifGPcolor
      \def\colorrgb#1{\color[rgb]{#1}}%
      \def\colorgray#1{\color[gray]{#1}}%
      \expandafter\def\csname LTw\endcsname{\color{white}}%
      \expandafter\def\csname LTb\endcsname{\color{black}}%
      \expandafter\def\csname LTa\endcsname{\color{black}}%
      \expandafter\def\csname LT0\endcsname{\color[rgb]{1,0,0}}%
      \expandafter\def\csname LT1\endcsname{\color[rgb]{0,1,0}}%
      \expandafter\def\csname LT2\endcsname{\color[rgb]{0,0,1}}%
      \expandafter\def\csname LT3\endcsname{\color[rgb]{1,0,1}}%
      \expandafter\def\csname LT4\endcsname{\color[rgb]{0,1,1}}%
      \expandafter\def\csname LT5\endcsname{\color[rgb]{1,1,0}}%
      \expandafter\def\csname LT6\endcsname{\color[rgb]{0,0,0}}%
      \expandafter\def\csname LT7\endcsname{\color[rgb]{1,0.3,0}}%
      \expandafter\def\csname LT8\endcsname{\color[rgb]{0.5,0.5,0.5}}%
    \else
      \def\colorrgb#1{\color{black}}%
      \def\colorgray#1{\color[gray]{#1}}%
      \expandafter\def\csname LTw\endcsname{\color{white}}%
      \expandafter\def\csname LTb\endcsname{\color{black}}%
      \expandafter\def\csname LTa\endcsname{\color{black}}%
      \expandafter\def\csname LT0\endcsname{\color{black}}%
      \expandafter\def\csname LT1\endcsname{\color{black}}%
      \expandafter\def\csname LT2\endcsname{\color{black}}%
      \expandafter\def\csname LT3\endcsname{\color{black}}%
      \expandafter\def\csname LT4\endcsname{\color{black}}%
      \expandafter\def\csname LT5\endcsname{\color{black}}%
      \expandafter\def\csname LT6\endcsname{\color{black}}%
      \expandafter\def\csname LT7\endcsname{\color{black}}%
      \expandafter\def\csname LT8\endcsname{\color{black}}%
    \fi
  \fi
    \setlength{\unitlength}{0.0500bp}%
    \ifx\gptboxheight\undefined%
      \newlength{\gptboxheight}%
      \newlength{\gptboxwidth}%
      \newsavebox{\gptboxtext}%
    \fi%
    \setlength{\fboxrule}{0.5pt}%
    \setlength{\fboxsep}{1pt}%
\begin{picture}(5102.00,3400.00)%
    \gplgaddtomacro\gplbacktext{%
      \csname LTb\endcsname
      \put(814,929){\makebox(0,0)[r]{\strut{}$1.4$}}%
      \put(814,1604){\makebox(0,0)[r]{\strut{}$1.7$}}%
      \put(814,2279){\makebox(0,0)[r]{\strut{}$2$}}%
      \put(814,2954){\makebox(0,0)[r]{\strut{}$2.3$}}%
      \put(946,484){\makebox(0,0){\strut{}$0.055$}}%
      \put(1669,484){\makebox(0,0){\strut{}$0.06$}}%
      \put(2392,484){\makebox(0,0){\strut{}$0.065$}}%
      \put(3115,484){\makebox(0,0){\strut{}$0.07$}}%
      \put(3838,484){\makebox(0,0){\strut{}$0.075$}}%
      \put(4560,484){\makebox(0,0){\strut{}$0.08$}}%
      \put(1134,2932){\makebox(0,0)[l]{\strut{}$\chi^2/{\mathrm{d.o.f}}=1.05$}}%
    }%
    \gplgaddtomacro\gplfronttext{%
      \csname LTb\endcsname
      \put(209,1941){\rotatebox{-270}{\makebox(0,0){\strut{}$B_4(u_c,\kappa, N_s)$}}}%
      \put(2825,154){\makebox(0,0){\strut{}$\kappa_c$}}%
      \csname LTb\endcsname
      \put(3604,1337){\makebox(0,0)[l]{\strut{}16x16x16}}%
      \csname LTb\endcsname
      \put(3604,1117){\makebox(0,0)[l]{\strut{}20x20x20}}%
      \csname LTb\endcsname
      \put(3604,897){\makebox(0,0)[l]{\strut{}24x24x24}}%
    }%
    \gplbacktext
    \put(0,0){\includegraphics[width={255.10bp},height={170.00bp}]{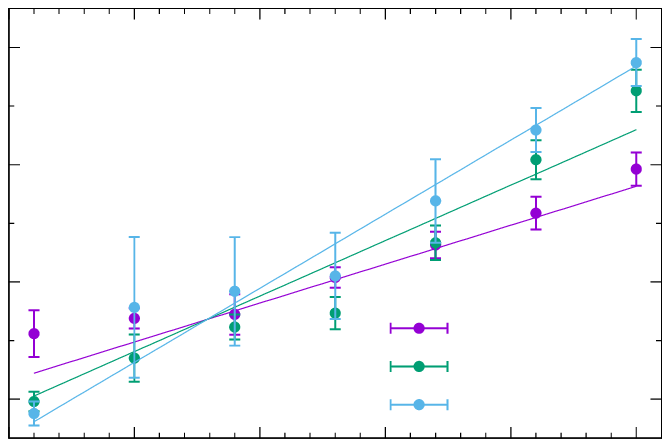}}%
    \gplfronttext
  \end{picture}%
\endgroup
}
	\caption{$S_{\text{eff}}^g+S_0+S_2, N_f=2, N_t=4$}
\end{subfigure}%
\begin{subfigure}{.5\textwidth}
\centering
	\scalebox{.9}{
\begingroup
  \makeatletter
  \providecommand\color[2][]{%
    \GenericError{(gnuplot) \space\space\space\@spaces}{%
      Package color not loaded in conjunction with
      terminal option `colourtext'%
    }{See the gnuplot documentation for explanation.%
    }{Either use 'blacktext' in gnuplot or load the package
      color.sty in LaTeX.}%
    \renewcommand\color[2][]{}%
  }%
  \providecommand\includegraphics[2][]{%
    \GenericError{(gnuplot) \space\space\space\@spaces}{%
      Package graphicx or graphics not loaded%
    }{See the gnuplot documentation for explanation.%
    }{The gnuplot epslatex terminal needs graphicx.sty or graphics.sty.}%
    \renewcommand\includegraphics[2][]{}%
  }%
  \providecommand\rotatebox[2]{#2}%
  \@ifundefined{ifGPcolor}{%
    \newif\ifGPcolor
    \GPcolortrue
  }{}%
  \@ifundefined{ifGPblacktext}{%
    \newif\ifGPblacktext
    \GPblacktexttrue
  }{}%
  \let\gplgaddtomacro\g@addto@macro
  \gdef\gplbacktext{}%
  \gdef\gplfronttext{}%
  \makeatother
  \ifGPblacktext
    \def\colorrgb#1{}%
    \def\colorgray#1{}%
  \else
    \ifGPcolor
      \def\colorrgb#1{\color[rgb]{#1}}%
      \def\colorgray#1{\color[gray]{#1}}%
      \expandafter\def\csname LTw\endcsname{\color{white}}%
      \expandafter\def\csname LTb\endcsname{\color{black}}%
      \expandafter\def\csname LTa\endcsname{\color{black}}%
      \expandafter\def\csname LT0\endcsname{\color[rgb]{1,0,0}}%
      \expandafter\def\csname LT1\endcsname{\color[rgb]{0,1,0}}%
      \expandafter\def\csname LT2\endcsname{\color[rgb]{0,0,1}}%
      \expandafter\def\csname LT3\endcsname{\color[rgb]{1,0,1}}%
      \expandafter\def\csname LT4\endcsname{\color[rgb]{0,1,1}}%
      \expandafter\def\csname LT5\endcsname{\color[rgb]{1,1,0}}%
      \expandafter\def\csname LT6\endcsname{\color[rgb]{0,0,0}}%
      \expandafter\def\csname LT7\endcsname{\color[rgb]{1,0.3,0}}%
      \expandafter\def\csname LT8\endcsname{\color[rgb]{0.5,0.5,0.5}}%
    \else
      \def\colorrgb#1{\color{black}}%
      \def\colorgray#1{\color[gray]{#1}}%
      \expandafter\def\csname LTw\endcsname{\color{white}}%
      \expandafter\def\csname LTb\endcsname{\color{black}}%
      \expandafter\def\csname LTa\endcsname{\color{black}}%
      \expandafter\def\csname LT0\endcsname{\color{black}}%
      \expandafter\def\csname LT1\endcsname{\color{black}}%
      \expandafter\def\csname LT2\endcsname{\color{black}}%
      \expandafter\def\csname LT3\endcsname{\color{black}}%
      \expandafter\def\csname LT4\endcsname{\color{black}}%
      \expandafter\def\csname LT5\endcsname{\color{black}}%
      \expandafter\def\csname LT6\endcsname{\color{black}}%
      \expandafter\def\csname LT7\endcsname{\color{black}}%
      \expandafter\def\csname LT8\endcsname{\color{black}}%
    \fi
  \fi
    \setlength{\unitlength}{0.0500bp}%
    \ifx\gptboxheight\undefined%
      \newlength{\gptboxheight}%
      \newlength{\gptboxwidth}%
      \newsavebox{\gptboxtext}%
    \fi%
    \setlength{\fboxrule}{0.5pt}%
    \setlength{\fboxsep}{1pt}%
\begin{picture}(5102.00,3400.00)%
    \gplgaddtomacro\gplbacktext{%
      \csname LTb\endcsname
      \put(550,704){\makebox(0,0)[r]{\strut{}$0$}}%
      \put(550,1529){\makebox(0,0)[r]{\strut{}$1$}}%
      \put(550,2354){\makebox(0,0)[r]{\strut{}$2$}}%
      \put(550,3179){\makebox(0,0)[r]{\strut{}$3$}}%
      \put(778,484){\makebox(0,0){\strut{}$0.04$}}%
      \put(1736,484){\makebox(0,0){\strut{}$0.05$}}%
      \put(2694,484){\makebox(0,0){\strut{}$0.06$}}%
      \put(3651,484){\makebox(0,0){\strut{}$0.07$}}%
      \put(4609,484){\makebox(0,0){\strut{}$0.08$}}%
      \put(883,2932){\makebox(0,0)[l]{\strut{}$\chi^2/\mathrm{d.o.f}=1.13$}}%
    }%
    \gplgaddtomacro\gplfronttext{%
      \csname LTb\endcsname
      \put(209,1941){\rotatebox{-270}{\makebox(0,0){\strut{}$B_4(u_c,\kappa, N_s)$}}}%
      \put(2693,154){\makebox(0,0){\strut{}$\kappa_c$}}%
      \csname LTb\endcsname
      \put(3486,1337){\makebox(0,0)[l]{\strut{}12x12x12}}%
      \csname LTb\endcsname
      \put(3486,1117){\makebox(0,0)[l]{\strut{}16x16x16}}%
      \csname LTb\endcsname
      \put(3486,897){\makebox(0,0)[l]{\strut{}20x20x20}}%
    }%
    \gplbacktext
    \put(0,0){\includegraphics[width={255.10bp},height={170.00bp}]{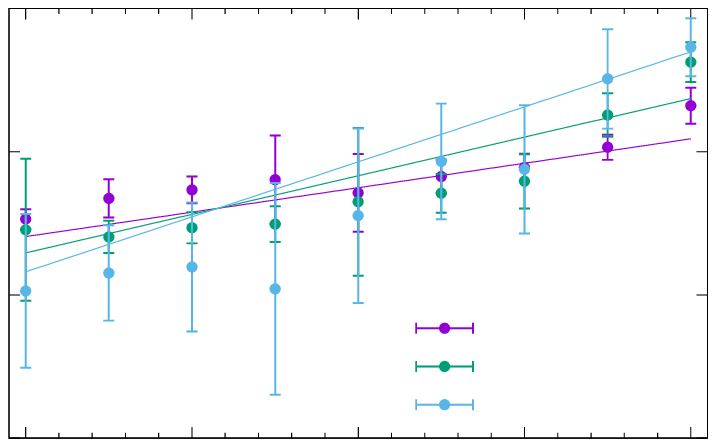}}%
    \gplfronttext
  \end{picture}%
\endgroup
}
	\caption{$S_{\text{eff}}^g+S_0+S_2+S_4, N_f=1, N_t=4$}
\end{subfigure}
\caption{Example plots of the kurtosis for different orders of the effective theory.} 
\label{fig:collapse_plot}
\end{center}
\end {figure}

Next, we find the location of the critical end point on the phase boundary. 
We plot the minimum values of $B_4$ against the couplings $h,\kappa$ for several volumes, see figure \ref{fig:collapse_plot}, 
and perform a linear fit to
equation~\eqref{eq:pseudoline_fit}. The $\chi^2/\text{d.o.f}$ values of the four fits are between $0.99-1.49$, indicating good fit qualities. 
With the same analysis, we estimate the critical values $\kappa_c$ also for the $\kappa^2$-action at $N_t=6$ and for 
the $\kappa^4$-action at $N_t=4$. All results are summarized in table \ref{table:critical_kappa_summary}. One observes that the value of $\kappa_c$ for $N_f=2$ of the same order effective theory is smaller than those of $N_f=1$. This is because the explicit 
center symmetry breaking is stronger with more fermion fields. 
The same argument can also be used to explain the decrease of $\kappa_c$ as the order in the hopping expansion in the effective theory increases. 

\begin{table}[t]
  \begin{center}
    \begin{tabular}{clcccccc}
    \toprule
   & & \multicolumn{2}{c}{eff.~theories (preliminary)} & \multicolumn{2}{c}{hopping expanded-QCD \cite{Ejiri:2019csa}} & \multicolumn{1}{c}{Full QCD \cite{Cuteri:2020yke}}\\
    \cmidrule(lr){3-4}\cmidrule(lr){5-6}\cmidrule(lr){7-7}
   $N_t$ & \text{action}  & $N_f=1$ & $N_f=2$ & $N_f=1$  & $N_f=2$  & $N_f=2$\\
      \hline
   &  $S_{\text{eff}}^g+S_0$ & 0.0810(4) & - & 0.0783(12) & 0.0658(10) & - \\
 $4$  &  $S_{\text{eff}}^g+S_0+S_2$ & 0.0756(6) & 0.0629(4) & 0.0753(11) & 0.0640(10)& -\\
   &   $S_{\text{eff}}^g+S_0+S_2+S_4$  & 0.0515(16) & 0.0443(34) & - & - & - \\
   $6$  &   $S_{\text{eff}}^g+S_0+S_2$  & 0.1319(6) & 0.1210(5) & 0.1326(21) & 0.1202(19) & 0.0877(9)\\
      \bottomrule
    \end{tabular}
    \caption{Comparison of the $\kappa_c$-values for the deconfinement critical point obtained by different approximations 
    to lattice QCD with Wilson quarks.}
    \label{table:critical_kappa_summary}
  \end{center}
\end{table}

\section{Conclusions}

In table~\ref{table:critical_kappa_summary}, we also compare our results with those obtained from $4$-dimensional QCD simulations
in the heavy quark region, in one case with a hopping expanded fermion determinant, in the other case with no approximations. 
We see that the phase structure of the $4$-dimensional full QCD is reproduced by the effective theories on a semi-quantitative
level, so that their application to the cold and dense regime can be trusted.
Regarding quantitative accuracy, the comparison with either hopping expanded or full QCD allows for detailed
conclusions regarding the strong coupling and hopping expansions separately: the 3D effective theory agrees almost quantitatively
with the hopping expanded 4D QCD, while both exhibit larger differences with full QCD as $N_t$ grows.
This means that the character expansion shows good convergence behavior and is sufficient for thermodynamical applications 
up to $N_t=6$, while higher order corrections are necessary in the hopping expansion already at $N_t=6$.   
\section*{Acknowledgments}
We acknowledge support by the Deutsche Forschungsgemeinschaft (DFG) through the grant
CRC-TR 211 “Strong-interaction matter under extreme conditions”.
\bibliography{./refs}

\end{document}